\documentclass{aa}
\usepackage{graphicx}
\usepackage{txfonts}
\usepackage{xcolor}
\usepackage{amsmath} 
\newcommand{\g}{$\gamma$}

\begin{document} 

   \title{The magnetic field structure in CTA~102 from high-resolution mm-VLBI observations during the flaring state in 2016-2017}


   \author{Carolina Casadio
          \inst{1}\fnmsep\thanks{\email{casadio@mpifr-bonn.mpg.de}}
          \and
          Alan P. Marscher\inst{2}
          \and
          Svetlana G. Jorstad \inst{2,3} 
          \and
          Dmitry A. Blinov \inst{4,3} 
          \and
          Nicholas R. MacDonald\inst{1}
          \and
          Thomas P. Krichbaum\inst{1}
          \and
          Biagina Boccardi\inst{1,5}
          \and
          Efthalia Traianou\inst{1}
          \and
          Jos\'e L. G\'omez\inst{6}
          \and
          Iv\'an Agudo\inst{6}
          \and
          Bong-Won Sohn\inst{7}
          \and
          Michael Bremer\inst{8}
          \and
          Jeffrey Hodgson\inst{7}
          \and
          Juha Kallunki\inst{9}
          \and
          Jae-Young Kim\inst{1}
          \and
          Karen E. Williamson\inst{2}
          \and
          J. Anton Zensus\inst{1}
          }

   \institute{Max-Planck-Institut f\"ur Radioastronomie, Auf dem   H\"ugel, 69, D-53121 Bonn, Germany\\
              \email{casadio@mpifr-bonn.mpg.de}
         \and
             Institute for Astrophysical Research, Boston University, Boston, MA 02215, USA 
             \and 
          Astronomical Institute, St. Petersburg State University, St. Petersburg 199034, Russia
         \and
         University of Crete, Heraklion, Greece
         \and
         INAF – OAS Bologna, Area della Ricerca CNR, Via Gobetti 101, I-40129 Bologna
         \and
          Instituto de Astrof\'{\i}sica de Andaluc\'{\i}a, CSIC, Apartado 3004, E-18080 Granada, Spain
          \and
          Korea Astronomy and Space Science Institute, 776 Daedeok-daero, Yuseong-gu, Daejeon 34055, Korea
          \and
          Institut de Radio Astronomie Millim{\'e}trique, 300 rue de la Piscine, 38406 Saint Martin d'H{\'e}res, France
          \and
          Aalto University Mets\"ahovi Radio Observatory, Mets\"ahovintie 114, 02540 Kylm\"al\"a, Finland
             }

   \date{Received 26 October 2018 / Accepted 1 December 2018}

 
  \abstract
   {Investigating the magnetic field structure in the innermost regions of relativistic jets is fundamental to understanding the crucial physical processes giving rise to jet formation, as well as to their extraordinary radiation output up to \g-ray energies.}
   {We study the magnetic field structure of the quasar CTA~102 with 3 and 7 mm VLBI polarimetric observations, reaching an unprecedented resolution ($\sim$50 $\mu$as). We also investigate the variability and physical processes occurring in the source during the observing period, which coincides with a very active state of the source over the entire electromagnetic spectrum.}
   {We perform the Faraday rotation analysis using 3 and 7 mm data and we compare the obtained rotation measure (RM) map with the polarization evolution in 7 mm VLBA images. We study the kinematics and variability at 7 mm and infer the physical parameters associated with variability. From the analysis of \g-ray and X-ray data, we compute a minimum Doppler factor value required to explain the observed high-energy emission.}
   {Faraday rotation analysis shows a gradient in RM with a maximum value of $\sim$6$\times$10$^{4}$ rad/m$^{2}$ and intrinsic electric vector position angles (EVPAs) oriented around the centroid of the core, suggesting the presence of large-scale helical magnetic fields. Such a magnetic field structure is also visible in 7 mm images when a new superluminal component is crossing the core region. The 7 mm EVPA orientation is different when the component is exiting the core or crossing a stationary feature at $\sim$0.1 mas. The interaction between the superluminal component and a recollimation shock at $\sim$0.1 mas could have triggered the multi-wavelength flares. The variability Doppler factor associated with such an interaction is large enough to explain the high-energy emission and the remarkable optical flare occurred very close in time.}
   {}

   \keywords{ active galactic nuclei--
               mm-VLBI  --
                rotation measure 
               }
\titlerunning{CTA~102 from High Resolution mm-VLBI}
\maketitle
%

\section{Introduction}

Collimated outflows are launched from the centers of powerful active galactic nuclei (AGNs), and propagate at relativistic speeds often far beyond the host galaxy. 
Helical magnetic fields anchored in either the ergosphere of a spinning supermassive black hole \citep{Blandford:1977ys} or the accretion disk surrounding it \citep{Blandford:1982fr, Zamaninasab:2014zr} are thought to collimate and power these relativistic outflows.
An observational signature of helical magnetic fields is a Faraday rotation gradient across the jet width due to the line-of-sight component of the magnetic field changing 
direction \citep[e.g.,][]{Laing:1981qd}. 

One of the objects in which a large-scale rotation measure (RM) gradient has been observed is the 
flat-spectrum radio quasar (FSRQ) CTA~102. \cite{Hovatta:2012fk} reported a significant transverse RM gradient at about 7 mas from the core region at 15 GHz.  
Hints that a helical magnetic field also exists in the innermost regions of the jets come from polarization observations at optical frequencies. The optical electric
vector position angles (EVPAs) have been observed rotating in coincidence with a multi-wavelength flaring event that occurred in 2012 and that also coincided with the ejection of a new superluminal component from the radio core  \citep{Larionov:2013uq, Casadio:2015fj}. 
Evidence for a physical connection between EVPA rotations and \g-ray flares has been found in a number of
other sources \citep[e.g.,][]{2010ApJ...710L.126M}. This connection cannot be entirely attributed to a random walk process \citep{2015MNRAS.453.1669B}.
Therefore, it is important to observe sources during outbursts in order to study the 
correlated variability at the different frequencies as well as to investigate changes in polarized emission. Moreover, in many blazars, including CTA~102, \g-ray 
outbursts seem to be triggered by the passage of traveling component(s) through 
the radio core. If the radio core is a standing shock and the traveling component(s) is(are) a perturbation(s) in the jet flow, we expect the magnetic field parallel to the shock front to be amplified during the passage of each component through the core \citep{1985ApJ...298..114M}. 
This would also result in the reordering of the local magnetic field which in turn would produce an increase in the observed polarized radio emission.  

We designed a monitoring program (PI: A. Marscher) consisting of Global millimeter(mm)-VLBI Array (GMVA) observations at 
3mm (86 GHz) of a sample of \g-ray bright blazars in support of the VLBA-BU-BLAZAR program\footnote{http://www.bu.edu/blazars/research.html}, which is a monthly monitoring with the VLBA at 7 mm (43 GHz) of 37 blazars and radio galaxies (e.g., Jorstad et al., 2017).  
The higher resolution ($\sim$50 $\mu$as) and lower opacity at 86 GHz (in comparison to 43 GHz) allow us to probe deeper into the innermost regions of the jet and to investigate the structural changes and physical conditions within the jet also in connection with the \g-ray emission, which is thought to originate mainly in these regions. 
The GMVA observations are performed roughly every six months and the objects observed consist of the brightest sources (roughly half) of the VLBA-BU-BLAZAR sample \citep{2017ApJ...846...98J}.   

In this paper we present results from the GMVA program obtained for
the FSRQ CTA~102 ($z$=1.037). The source recently underwent a prolonged active 
phase displaying, from the end of 2016 until beginning of 2017, a series of 
flares from the mm to the \g-ray energy band. A very bright flare occurred in the optical at the end of December 2016, when the source became the brightest blazar ever detected at these wavelengths. Coincident with this large optical flare were a series of outbursts that occurred closely in time across the \g-ray, X-ray, UV, and 1~mm radio bands \citep[e.g.,][]{2017Natur.552..374R,2018A&A...617A..59K}.

The 86 and 43 GHz VLBI polarimetric data presented in this work cover the observing period from May 2016 to March 2017. The data set used for the analysis and the methods adopted for the data calibration are described in Section~\ref{obs}. In Section~\ref{pol}, we analyze the polarized emission at 86 and 43 GHz and in Section~\ref{flares} we present our findings in connection with the multi-wavelength flares in 2016 - 2017.
 We adopt the cosmological values from the most recent {\it Planck} satellite results \citep{2016A&A...594A..13P}: $\Omega_{m}$= 0.3, $\Omega_{\Lambda}$= 0.7, and $H_{0}$ = 68 km s$^{-1}$ Mpc$^{-1}$. Assuming these values, and at the source redshift, 1 mas corresponds to a linear distance of 8.31 pc, and a proper motion of 1 mas yr$^{-1}$ corresponds to an apparent speed of 55.2$c$.
\section{Observations and data reduction}\label{obs}

\subsection{VLBI data analysis}

The GMVA polarimetric data we present here were obtained on 21 May and 30 September 2016, and 31 March 2017. The antennas joining the GMVA array are: 8$\times$VLBA stations (BR, FD, PT, LA, OV, KP, NL, MK), Green Bank Telescope (GB), Effelsberg (EB), Yebes (YS), Mets\"ahovi (MH), Onsala (ON), Pico Veleta (PV), Plateau de Bure (PdB), and the Korean VLBI Network (KVN) array. In Table~\ref{table:1} we report the antennas for which it was possible to detect fringes, in each of the three observing sessions. 
Data are recorded at a rate of 2 Gbps (512 MHz bandwidth) in dual-polarization mode at all stations, aside from Yebes which records in single polarization. Afterwards, during the correlation process, data were split into eight 32-MHz sub-bands per polarization (IFs).  

The {\it a priori} calibration (i.e., amplitude and phase calibration) of both 86 and 43 GHz data was performed following the usual procedure for high-frequency VLBI data reduction in Astronomical Image Processing System ({\tt AIPS}); see for example \cite{2017ApJ...846...98J}. 

Since the expected atmospheric coherence time at 86 GHz is very short ($\sim10-20$ sec), the phase stability and the accuracy of the amplitude calibration are more critical than at longer wavelengths. As reported in \cite{2017Galax...5...67C}, in order to check the reliability of the amplitude calibration, the final total flux density values have been compared with 3mm single-dish measurements from the IRAM 30m antenna obtained under the POLAMI program\footnote{See \cite{2018MNRAS.474.1427A,2018MNRAS.473.1850A} and \url{http://polami.iaa.es}}. In Table~\ref{table:2} we report the comparison of GMVA total flux densities and POLAMI (3mm) single dish measurements of near-in-time epochs. We notice that the GMVA flux is lower than the POLAMI measurement in all the three epochs. Since CTA~102 is a relatively compact source, we expected a better match between the two measurements; it is therefore possible that the mm-VLBI total flux is slightly underestimated. However, as our study is mainly focused on the analysis of the polarized emission of CTA~102, these differences in flux density do not have an appreciable effect on the results of our study.   
We also use the POLAMI program measurements for the calibration of the absolute EVPA orientation, with uncertainties in the EVPAs $\leq$ 5$^{\circ}$ \citep{2018MNRAS.474.1427A,2018MNRAS.473.1850A}.

\begin{table*}
\caption{Antennas participating in GMVA sessions}             
\label{table:1}      
\centering                          
\begin{tabular}{c c}        
\hline\hline                 
Epoch & Antennas  \\    
\hline                        
 21 May 2016 & VLBA, EB, ON, KVN \\      
 30 Sep 2016 & VLBA (- MK), EB, ON, YS, MH, GB, KVN  \\
 31 Mar 2017 & VLBA, EB, ON, YS, MH, PV, GB, KVN \\
\hline                                   
\end{tabular}
\end{table*}

\begin{table}
\caption{Single dish comparison. The near-in-time POLAMI epochs are: 14.05 and 14.06.2016, 20.09.2016, 31.03.2017.}             
\label{table:2}      
\centering                          
\begin{tabular}{c c c}        
\hline\hline                 
GMVA Epoch &  \multicolumn{2}{c}{Total Flux Density (Jy)}  \\
 & GMVA & POLAMI  \\    
\hline                        
 21.05.2016 & 2.74$\pm$0.28 & 4.82$\pm$0.18 \\  
 &  & 4.7$\pm$0.18 \\
 30.09.2016 &  2.23$\pm$0.22 & 6.44$\pm$0.24 \\
 31.03.2017 &  4.38$\pm$0.44 & 6.19$\pm$0.32 \\
\hline                                   
\end{tabular}
\end{table}

The calibration of instrumental polarization is another complex step of polarimetric data reduction. After the a priori calibration, the data are transferred into {\tt Difmap} to obtain the first total intensity image through a combination of CLEAN and self calibration. Afterwards, data are brought back to {\tt AIPS} for the polarization calibration which consists of correcting for the instrumental polarization ({$D$-terms}) and the EVPAs absolute orientation.
We tested two different methods for the calibration of instrumental polarization and used the most reliable one for the final calibration as described in Appendix~\ref{app:A}. 

To carry out an analysis of the jet kinematics and flux density variability, we fitted the visibilities in {\tt Difmap} with circular Gaussian components describing the brightness distribution. For each model-fit component, both at 3 and 7 mm, we obtain the flux density and position and we infer the relative uncertainties using the empirical relation in \cite{Casadio:2015fj}. Moreover, we added  an additional 10$\%,$ coming from the typical amplitude calibration errors, and a minimum positional error of 0.005 mas, corresponding to $\sim$ 1/5 of the observing beam, (in quadrature) to the uncertainties above, as in \cite{2017ApJ...846...98J}. 

We calculate the uncertainties on the degree of linear polarization ($\sigma_{m}$) using the error propagation theory:
\\
\begin{align}
\label{sigmap}
\sigma_{p}&=\frac{\sqrt{(Q~\sigma_{Q})^{2} + (U~\sigma_{U})^{2}}}{P}\\
\sigma_{m}&=\frac{1}{I}\sqrt{\sigma_{p}^{2} + \left(\frac{P}{I}\times\sigma_{I}\right)^{2}},
\label{sigmap2}
\end{align}
\\
where $\sigma_{Q}$ and $\sigma_{U}$ are the rms noise of Stokes $Q$ and $U$ images, P = $\sqrt{Q^{2} + U^{2}}$ is the linearly polarized flux density and $I$ is the flux density in Stokes $I$, both with the respective uncertainties to which we added (in quadrature) a calibration uncertainty of 10$\%$ \citep{2014A&A...571A..54L}. 
To the $\sigma_{m}$ obtained in Eq.~\ref{sigmap2} we add, in quadrature, $\sigma_{m,D}$, which is the feed calibration error we obtain from the analysis of the $D$-terms and is estimated using the following formula from \cite{1994ApJ...427..718R}: 
\\
\begin{align}
\label{sigmamD}
\sigma_{m,D}=\sigma_{D}(N_{a}N_{IF}N_{s})^{-1/2},
\end{align}
\\
where $\sigma_{D}$ is the standard deviation related to the weighted average $D$-term measurements (see Appendix~\ref{app:A}) and is of the order of 1 - 3$\%$, $N_{a}$ is the number of antennas, $N_{IF}$ the number of IFs, and $N_{s}$ is the number of sources we used to infer the weighted average $D$-terms data set. In the original formula \citep{1994ApJ...427..718R}, $N_{s}$ is the number of scans with independent parallactic angles, which in our case would be a larger quantity than the considered $N_{s}$, giving consequently even smaller $\sigma_{m,D}$. We obtained $\sigma_{m,D}\approx$ 0.1$\%$ and a final $\sigma_{m}$ for the three epochs of between 0.4 and 0.7$\%$.
From the comparison of the different methods tested for the calibration of instrumental polarization in Appendix~\ref{app:A}, we notice that an imperfect $D$-term calibration could cause substantial changes in the polarized image and consequently in the degree of linear polarization. 
In contrast, the EVPA orientation was fairly constant with all the different methods; therefore for the EVPA errors we considered only the uncertainties of the POLAMI measurements.

\subsection{X-ray and Gamma-ray data analysis}

For the acquisition of the \g-ray photon fluxes of CTA~102 we analyzed the publicly available {\em Fermi} Large Area Telescope (LAT) data \citep{Atwood2009}. Using the unbinned likelihood analysis of the {\em Fermi} analysis software package Science Tools v10r0p5 we processed the data in the $100\, {\rm MeV} \le E \le 100\, {\rm GeV}$ energy range. Within the $15\deg$ region of interest centered on the blazar we selected source class photons (evclass=128 and evtype=3). Photons with high satellite zenith angle ($\ge 90\deg$) were omitted in order to exclude the Earth limb background. The spatial model $gll\_iem\_v06$ was used to account for the diffuse emission from the Galaxy, while the isotropic spectral template $iso\_source\_v05$ was included in the fit for modeling of the extragalactic diffuse and residual instrumental backgrounds. We used $P8R2\_SOURCE\_V6$ as the instrument response function. In the background model we included all sources from the 3FGL \citep[the third {\em Fermi} Gamma-ray catalog,][]{Acero2015} located within $15\deg$ of the blazar. Spectral shapes of all targets except CTA~102 and photon fluxes of sources farther than $10\deg$ from the blazar were fixed to their values listed in the 3FGL. We used the test statistic value $TS=10$ as  the detection limit. This approximately corresponds to a $3\sigma$ detection level \citep{Nolan2012}. The photon flux was integrated within six-hour time bins in order to maximize both the resolution of the light curve and the number of detections in it.

We collected X-ray data within the time range of interest from the Swift archive. The X-ray Telescope (XRT) \citep{Burrows2004} data were taken in the energy range 0.3–10 keV in Photon Counting mode and processed with the standard HEAsoft package version 6.19 in the manner described in \cite{Williamson2014}. XSPEC software \citep{1996ASPC..101...17A} was used to fit the spectra by a single power-law model, with 
the neutral hydrogen column density equal to 5.04$\times$10$^{20}$cm$^{-2}$ \citep{1990ARA&A..28..215D}.  We employed the Monte Carlo method in XSPEC to determine the goodness of the fit and uncertainties for each spectrum. The uncertainties of measurements are given at the 90$\%$ confidence level.

\section{Polarization and rotation measure analysis}
\label{pol}

   \begin{figure*}
   \centering
   \includegraphics[width=0.7\textwidth]{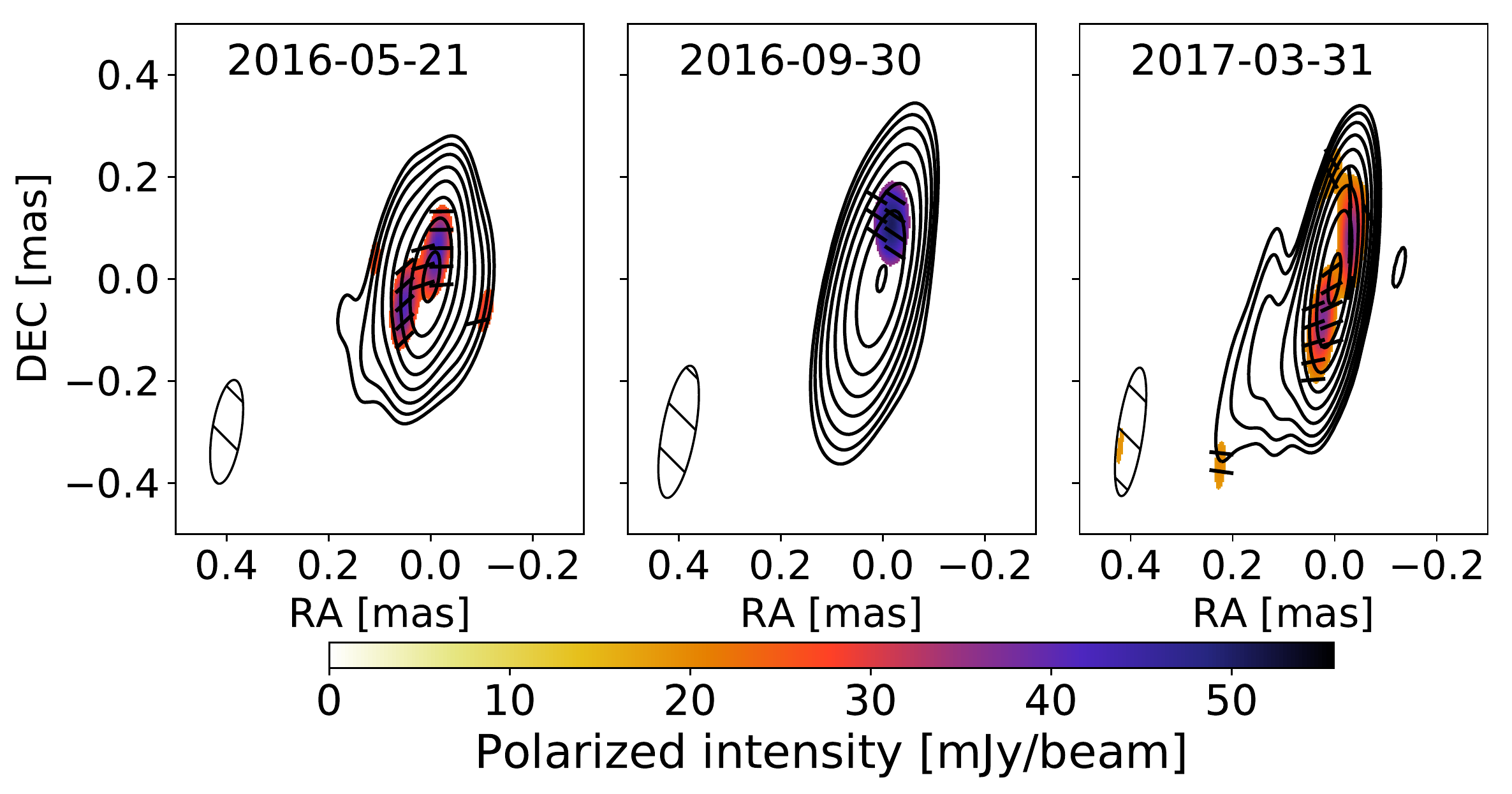}
   \caption{86 GHz GMVA polarimetric images of CTA~102. The restoring beams are 0.21$\times$0.06, 0.26$\times$0.06 and 0.25$\times$0.05 mas, respectively. Total intensity peaks are 2.0, 1.77, and 3.79 Jy/beam and contours are drawn at 0.5, 0.85, 1.53, 3.02, 5.96, 11.74, 23.15, 45.65, and 90 $\%$ of 3.79 Jy/beam.}
              \label{GMVA}%
    \end{figure*}

In Fig.~\ref{GMVA} we display the three GMVA polarimetric images covering the period from  May 2016 to March 2017. The source structure at 86 GHz in total intensity is relatively compact, while in linearly polarized intensity we can discern different features. 
In particular, it is interesting to notice the polarized emission downstream of the peak in total intensity, where the EVPAs have the same orientation as the jet direction for the May 2016 and March 2017 epochs. The maximum degree of polarization for the three epochs is comparable: between 8 and 14$\%$. 

In order to recover more polarized emission and lower the rms of the map, we also performed the stacking (average) of the three images in both total and linearly polarized intensity. For the linearly polarized intensity the stacking was done separately for the images of the Q and U Stokes parameters. The averaging in the image plane was performed after the alignment of the maps using the position of the peak in total intensity, which was coincident with the core position in all three epochs. Another method for the alignment of images makes use of the position of the VLBI core, which is determined  through the {\it modelfit} procedure in {\tt Difmap} package \citep[e.g.,][]{2017MNRAS.468.4992P}. We compared both methods and found the first method to give better images (higher dynamic range) in the case of 86 GHz data. This could be a consequence of the higher angular resolution at 86 GHz, which causes the position of the core model-fit component to vary substantially among epochs. 

In order to study the intrinsic magnetic field orientation we need to correct for the EVPA rotation introduced by the Faraday rotation. When the electromagnetic wave crosses a magnetized plasma, the polarization plane rotates due to a different propagation velocity of the left and right circularly polarized waves in which the linearly polarized radiation can be decomposed. The intrinsic polarization plane ($\chi_{0}$) is therefore rotated by a quantity, RM, which depends on the magnetic field (B) along our line of sight and the electron density (n$_{e}$) of the intervening plasma:
\\
\begin{align}
\chi=\chi_{0}+\frac{e^{3} \lambda^{2}}{8\pi^{2}\epsilon_{0}m^{2}c^{3}}\int{n_{e}\mathbf{B}\cdot dl}=\chi_{0}+RM\lambda^{2}
\label{eq1}
\end{align}
\\
where $e$ is the electron charge, $\epsilon_{0}$ the vacuum permittivity, $m$ the electron mass, and $c$ the speed of light. Given the linear dependence between the observed EVPAs and wavelength squared ($\lambda^{2}$) in Eq.~\ref{eq1}, the RM can be estimated from EVPA measurements at several frequencies.

\begin{figure}
   \centering
   \includegraphics[width=\hsize]{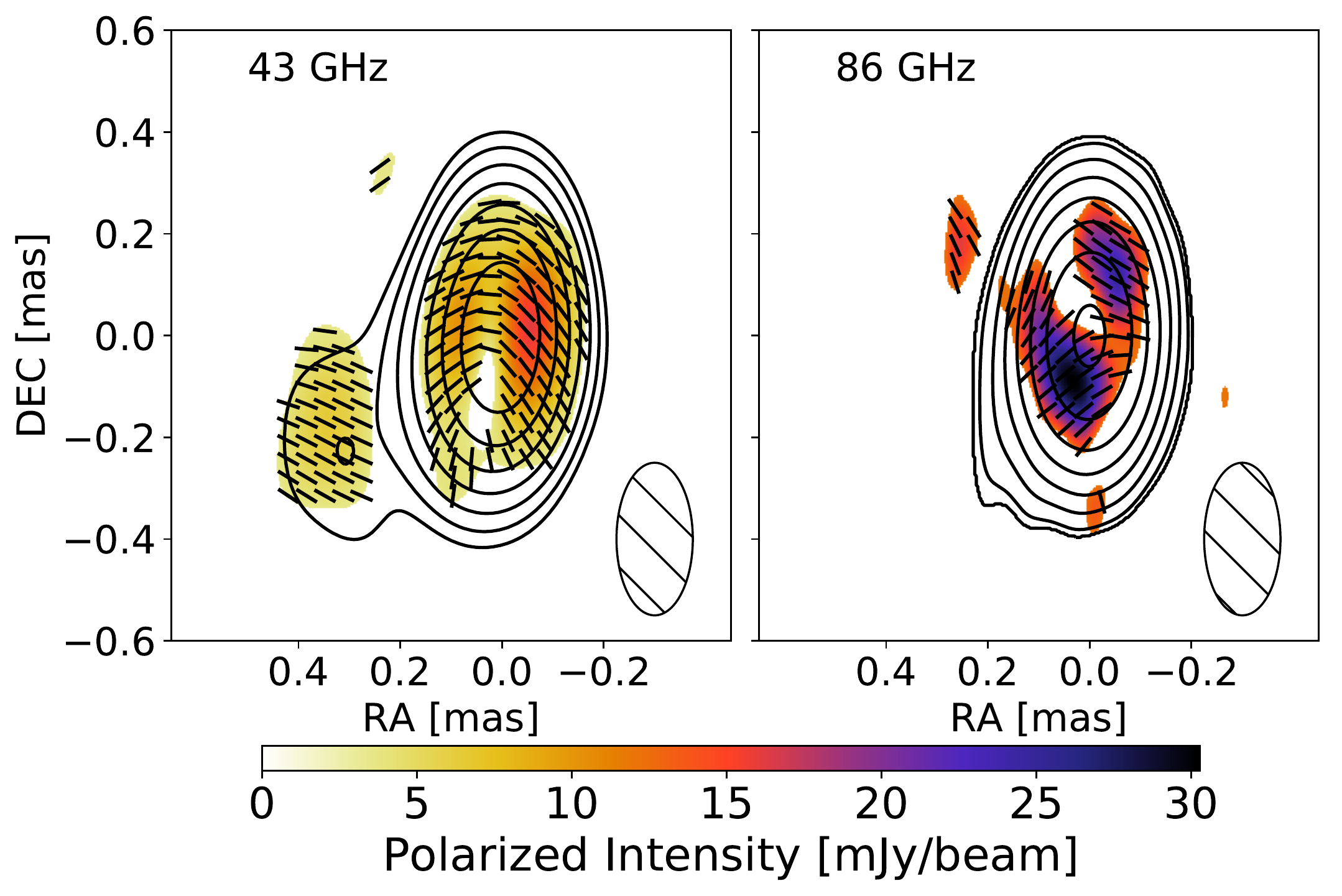}
   \caption{43 GHz VLBA (left) and 86 GMVA (right) stacked images. Black sticks represent EVPAs. The common restoring beam of 0.3$\times$0.15 mas is displayed in the bottom-right corner.}
              \label{concatenate}%
    \end{figure}

We collected 43 GHz data from the VLBA-BU-BLAZAR program from June 2016 to April 2017, covering a time range similar to that covered by our GMVA epochs. The resulting polarimetric images were convolved with the same restoring beam (0.3$\times$0.15 mas, 0$^{\circ}$); we did the same for the GMVA images. The resultant 43- and 86-GHz images have been stacked using the method described above. The two stacked images at 43- and 86-GHz, displayed in Fig.~\ref{concatenate}, were used to obtain the RM image of CTA~102 between these two frequencies.

To obtain the RM image we first aligned the two images. Since the source is very compact at these frequencies, it was impossible to identify common optically thin regions. Therefore, we aligned the two stacked maps using a cross-correlation algorithm based on the correlation of total intensity images \citep[e.g.,][]{Hovatta:2012fk, 2016ApJ...817...96G}. We obtained a shift to the southwest of 0.017 mas to align the 43 GHz image with respect to the 86 GHz one.  

\begin{figure}
   \centering
   \includegraphics[width=0.8\hsize]{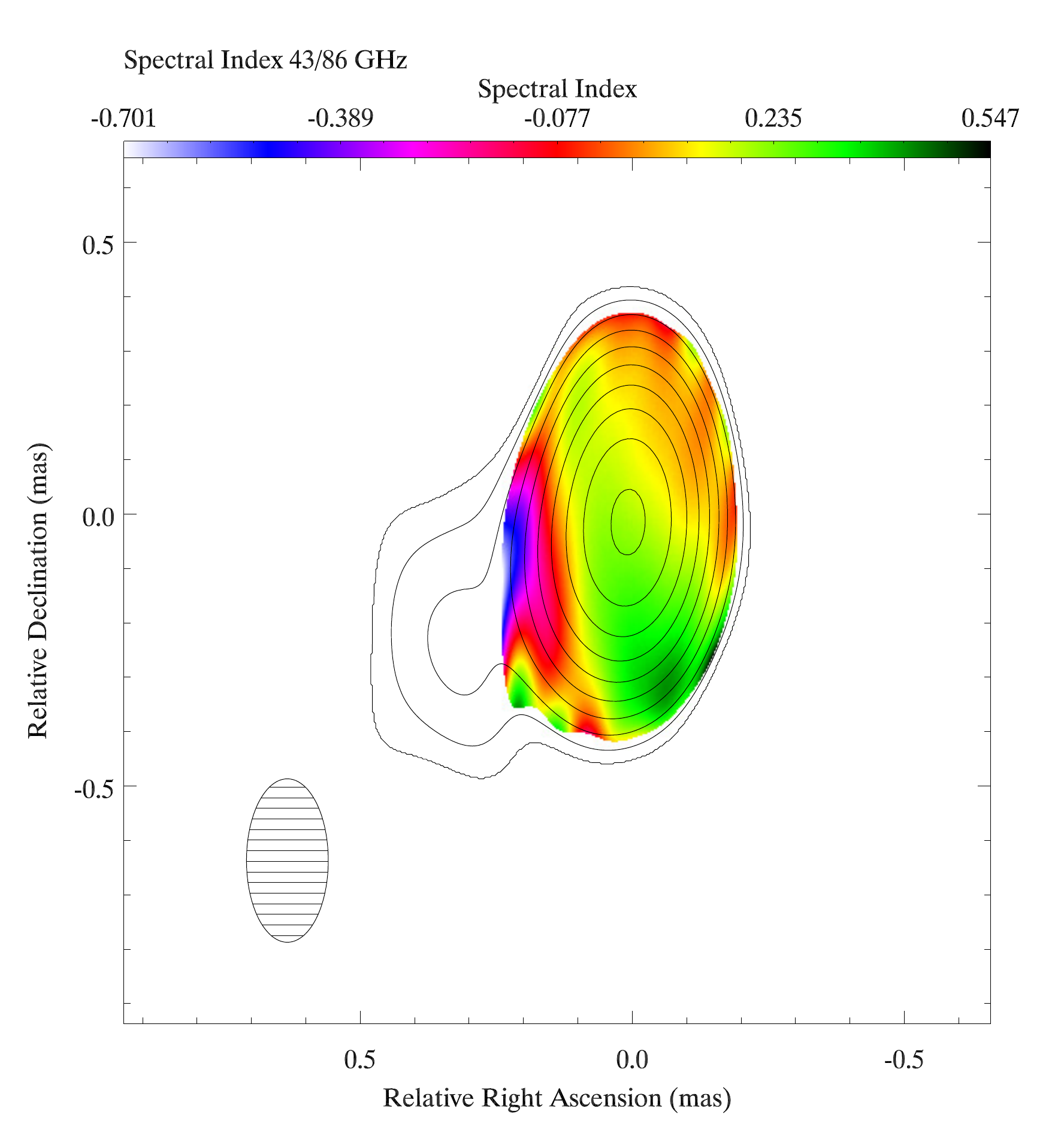}
   \caption{Spectral index image between the 43 GHz VLBA  and 86-GHz GMVA stacked images. Contours display the 43 GHz total intensity stacked map. The common restoring beam of 0.3$\times$0.15 mas is displayed on the left.}
              \label{sp4386}%
    \end{figure}

We also computed the spectral index (S$_\nu\propto\nu^\alpha$) map in order to check for optically thin/thick transitions in the core emitting region, where the EVPAs are expected to rotate by $\pi$/2 \citep{1970ranp.book.....P}. Figure~\ref{sp4386} shows the spectral index map between 86 GHz GMVA and 43 GHz VLBA stacked images. Most of the core region is optically thick, becoming optically thin at $\sim$0.15 mas downstream from the core. Since the polarized emission is limited to the core region, which is mainly optically thick, we did not apply any rotation to EVPAs associated with opacity.   

\begin{figure}
   \centering
   \includegraphics[width=\hsize]{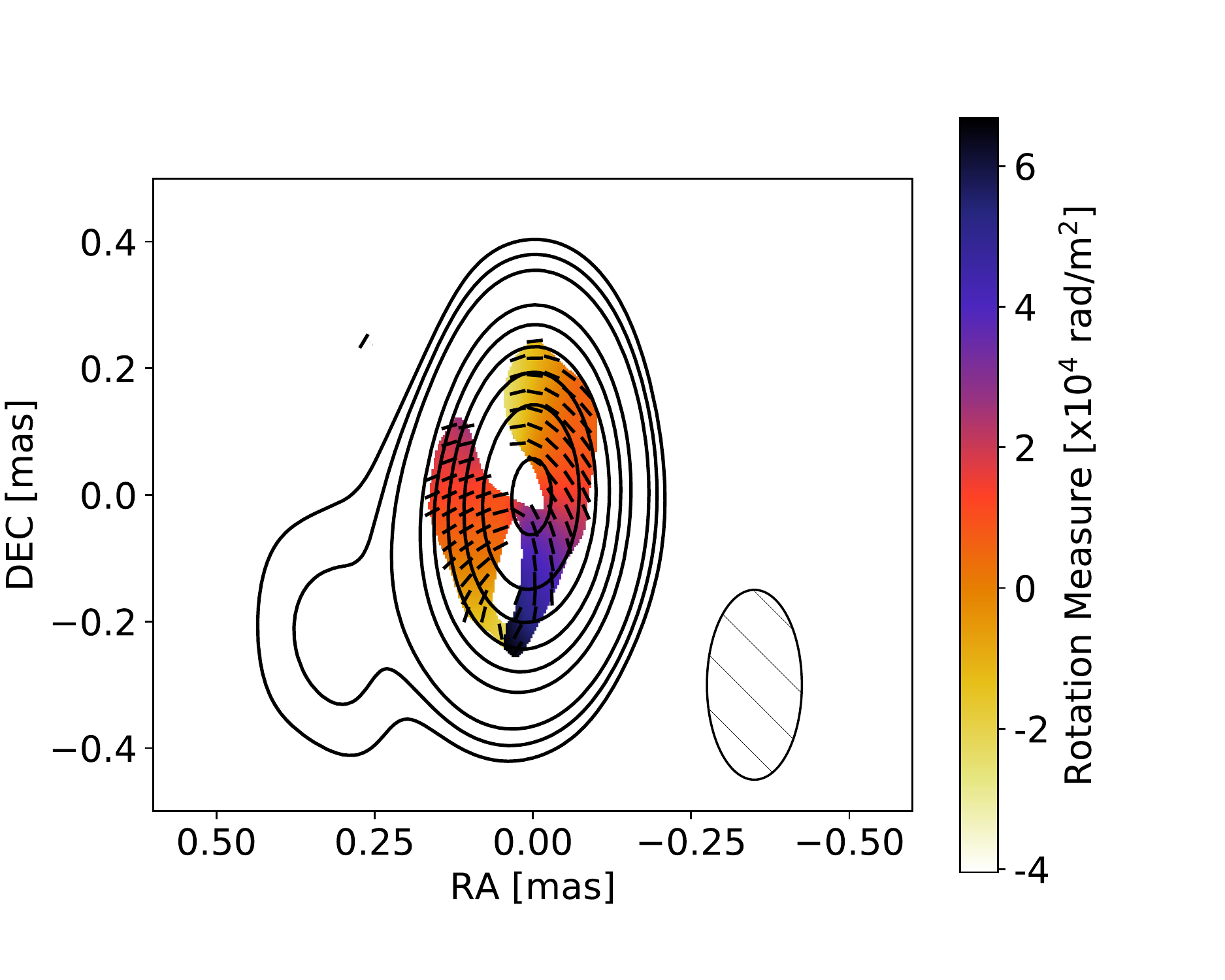}
   \caption{Rotation measure image using the 43 and 86 GHz stacked images. Colors show the RM and black sticks the intrinsic (Faraday-corrected) EVPAs. The restoring beam and contours are the same as in Fig.~\ref{sp4386}.}
              \label{RM}%
    \end{figure}

The RM map was obtained with the same approach described in \cite{2016ApJ...817...96G}. We made use of an IDL routine which estimates in each pixel the minimum RM value (i.e., the value that minimizes the $n\pi$ ambiguity in EVPAs). In Fig.~\ref{RM} we present the resultant RM map.


The Faraday rotation analysis between 43 and 86 GHz reveals RM values in the core region  ranging from $\sim$ $-$2$\times$10$^{4}$ to $\sim$6$\times$10$^{4}$ rad/m$^{2}$, in agreement with values reported for this source in previous studies \citep{2007AJ....134..799J,2018ApJ...860..112P}. It also reveals a gradient around the centroid of the core and a change of sign, clearly  visible from Fig.~\ref{RM}. The intrinsic EVPAs (i.e., the EVPAs corrected for Faraday rotation), represented by black sticks in Fig.~\ref{RM}, seem to rotate around the central peak.    

The RM gradient, the change of sign, and the peculiar orientation of intrinsic EVPAs resemble the situation in BL~Lacertae, as recently found by \cite{2016ApJ...817...96G}. The authors, who performed the Faraday rotation analysis using 15 and 43 GHz VLBA data and 22 GHz {\it RadioAstron} data, associate their findings to the presence of a large-scale helical magnetic field. 

Relativistic magnetohydrodynamic (MHD) simulations including such large-scale helical magnetic fields in a Faraday rotating sheath surrounding the jet predict transverse gradient in RM \citep{2010ApJ...725..750B}. In addition, the observed EVPA orientation around the core center has been produced by simulations in which the jet is observed at very small viewing angles \citep{2011ApJ...737...42P}.
Hence we conclude that the core region in CTA~102 is threaded by a large-scale helical magnetic field.

\section{The connection with the multi-wavelength flares in 2016 - 2017}
\label{flares}

\subsection{The gamma-ray doppler factor}
   \begin{figure*}
   \centering
   \includegraphics[width=0.85\textwidth]{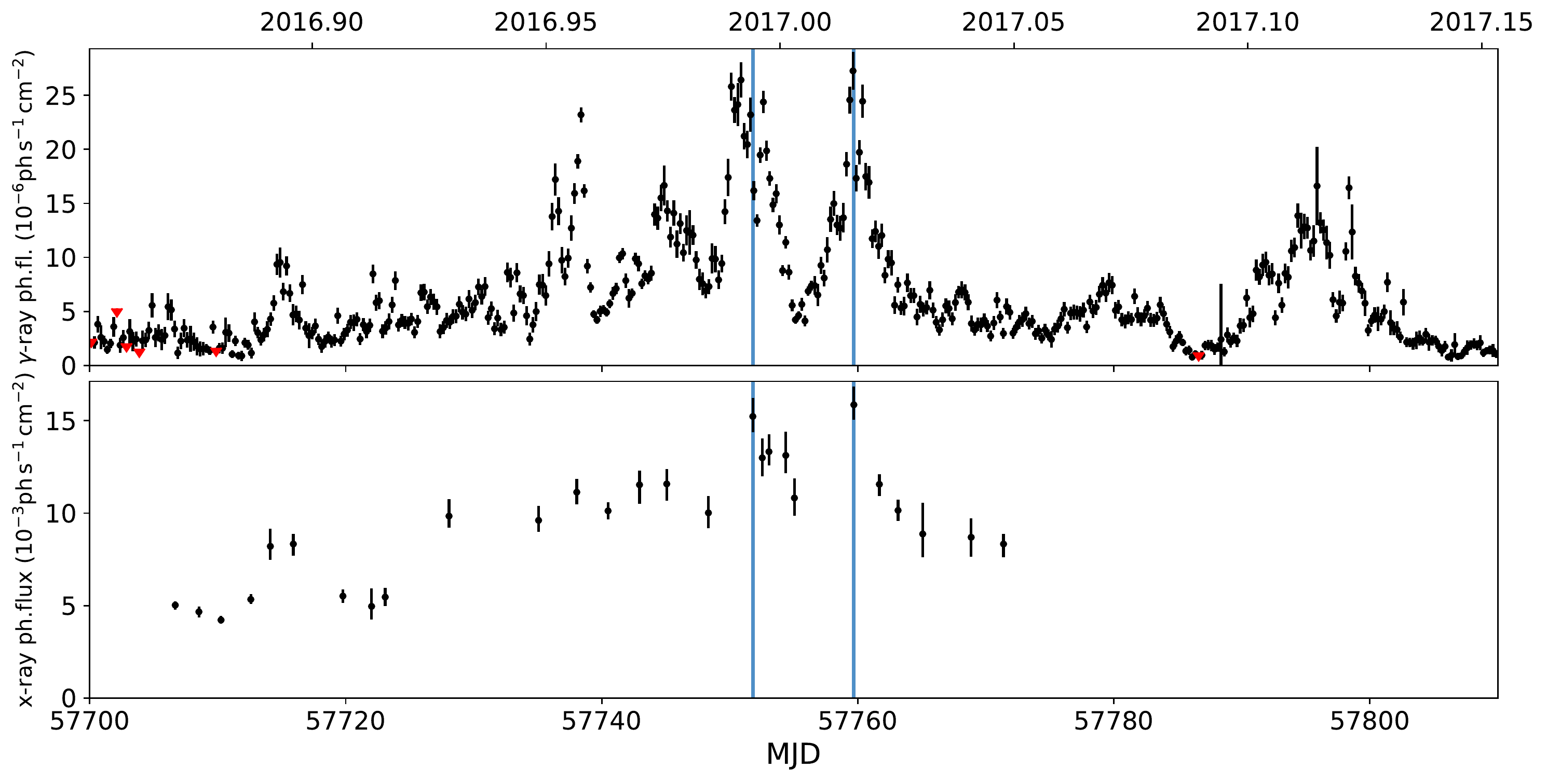}
   \caption{Photon flux curves during the flaring period in $\gamma$-rays (top) and X-rays (bottom). The red triangles represent upper limits of the photon flux.}
              \label{FigGamXray}%
    \end{figure*}
In Fig.~\ref{FigGamXray} we show \g-ray (top panel) and X-ray (bottom panel) light curves, confirming the high state(s) of activity at both frequencies during December 2016 - January 2017 ($\sim$57735 - 57760 MJD).
In both light curves we can separate two peaks: the first one around 57752 MJD (30 December) and the second 8 days later, around 57760 MJD.

Given the fact that we detect X-ray and \g-ray emission and the timing between the two frequencies suggests a common origin, we can use the method described in \cite{Mattox1993} to derive a lower limit for the Doppler factor ($\delta_{\gamma}$). This method is based on the maximum optical depth allowed in order to avoid pair-production absorption, given a plasma moving relativistically as in AGN jets. 
We then inferred $\delta_{\gamma}$ for the two events using equation (4) in \cite{Mattox1993} and the parameters reported in Table~\ref{table:3}.

\begin{table}
\caption{Physical parameters used for deriving $\delta_{min}$ from equation (4) in \cite{Mattox1993}}            
\label{table:3}      
\centering                         
\begin{tabular}{c c c c c c}        
\hline\hline                 
MJD & $\alpha$ & $h_{75}$ & $T_{5}$ & $F_{keV}$ & $E_{\gamma}$ \\
\hline                        
57754 & 0.17 & 0.9 & 0.58 & 3.22 & 100\\  
57760& 0.29 & 0.9 & 1.72 & 4.14 & 100\\
\hline                                   
\end{tabular}
\tablefoot{$\alpha$ is the X-ray spectral index, $h_{75}$ is derived from cosmological parameters in \cite{2016A&A...594A..13P}, $T_{5}$ is the variability time scale in X-ray in unit of 10$^{5}$ sec, $F_{keV}$ is the flux at 1 keV in unit of $\mu$Jy, and $E_{\gamma}$ is the highest $\gamma$-ray photon energy in GeV unit.}
\end{table}
The variability timescales during the X-ray events in 57752 and 57760 MJD are $\sim$16 and 48 hours, respectively, and these are the fastest significant variations we can derive from the X-ray light curve. Nevertheless these values could be underestimated, since a better sampling could give faster variability. 
The highest photon energy detected 
in the \g-rays during the flaring period is 364 $\mathrm{GeV}$ but since this value comes from only one photon we opted for a more conservative approach and used the value of 100 GeV, which is in agreement with the $\sim$98 GeV reported in \cite{2018ApJ...863..114G}.

For the two high-energy events we obtained $\delta_{\gamma}\gtrsim$ 17 during the first outburst (57752 MJD) and $\delta_{\gamma}\gtrsim$ 15 during the second (57760 MJD). Given the similarity of the two events in both the X-ray and \g-ray bands, similar limiting values for the Doppler factor were expected.

\subsection{The Kinematics at 43 GHz and the variability Doppler factor}
\begin{figure}
   \centering
   \includegraphics[width=\hsize]{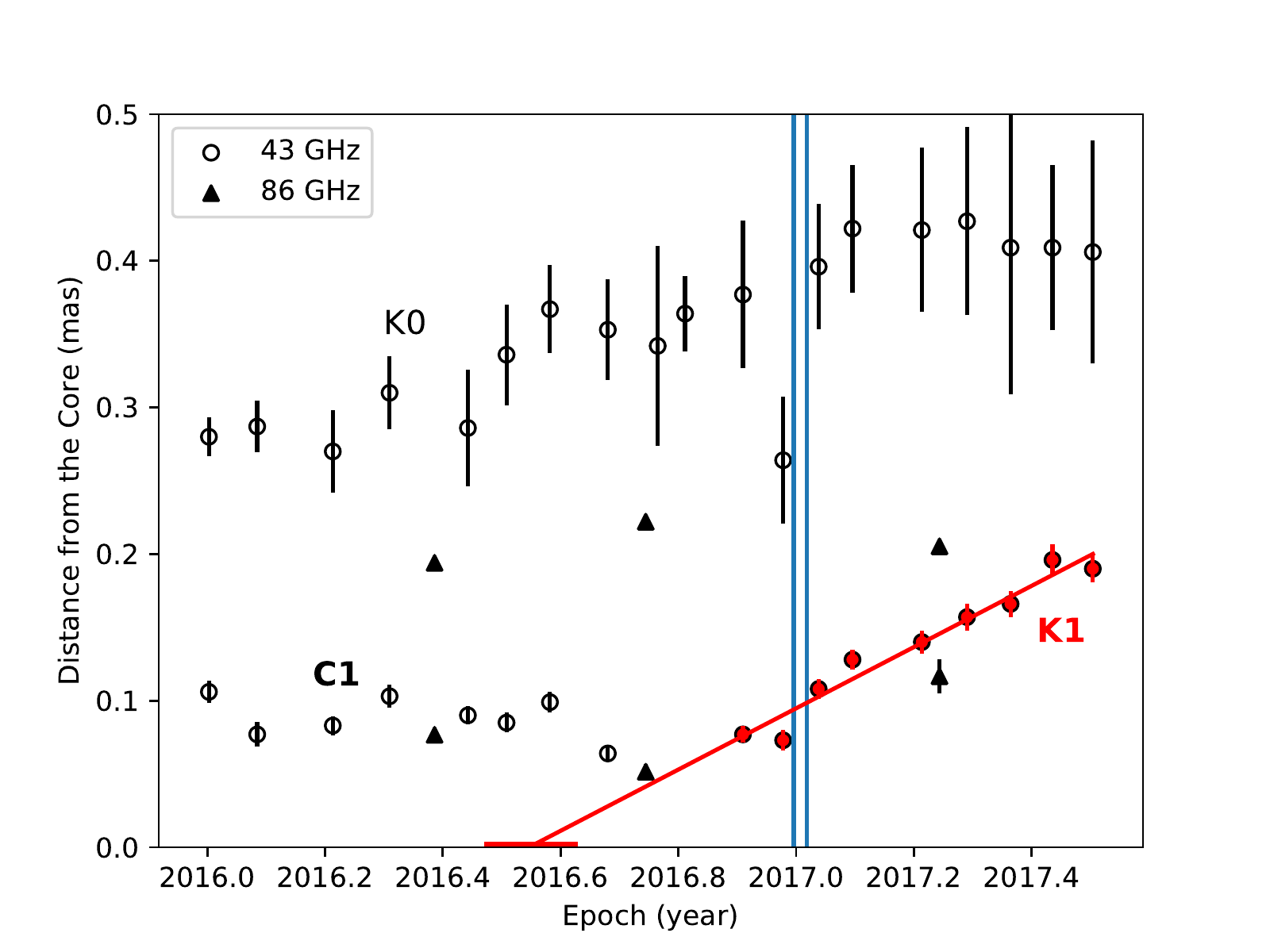}
   \caption{Distance from the core vs. time of the 43 and 86 GHz model-fit components. The red line is the linear fit of K1 positions plus the uncertainty associated to the ejection time of the component. Blue vertical lines mark the time of the two high-energy events discussed in Sect.~\ref{flares}.}
              \label{Kinematic43}%
    \end{figure}

\begin{figure}
   \centering
   \includegraphics[width=\hsize]{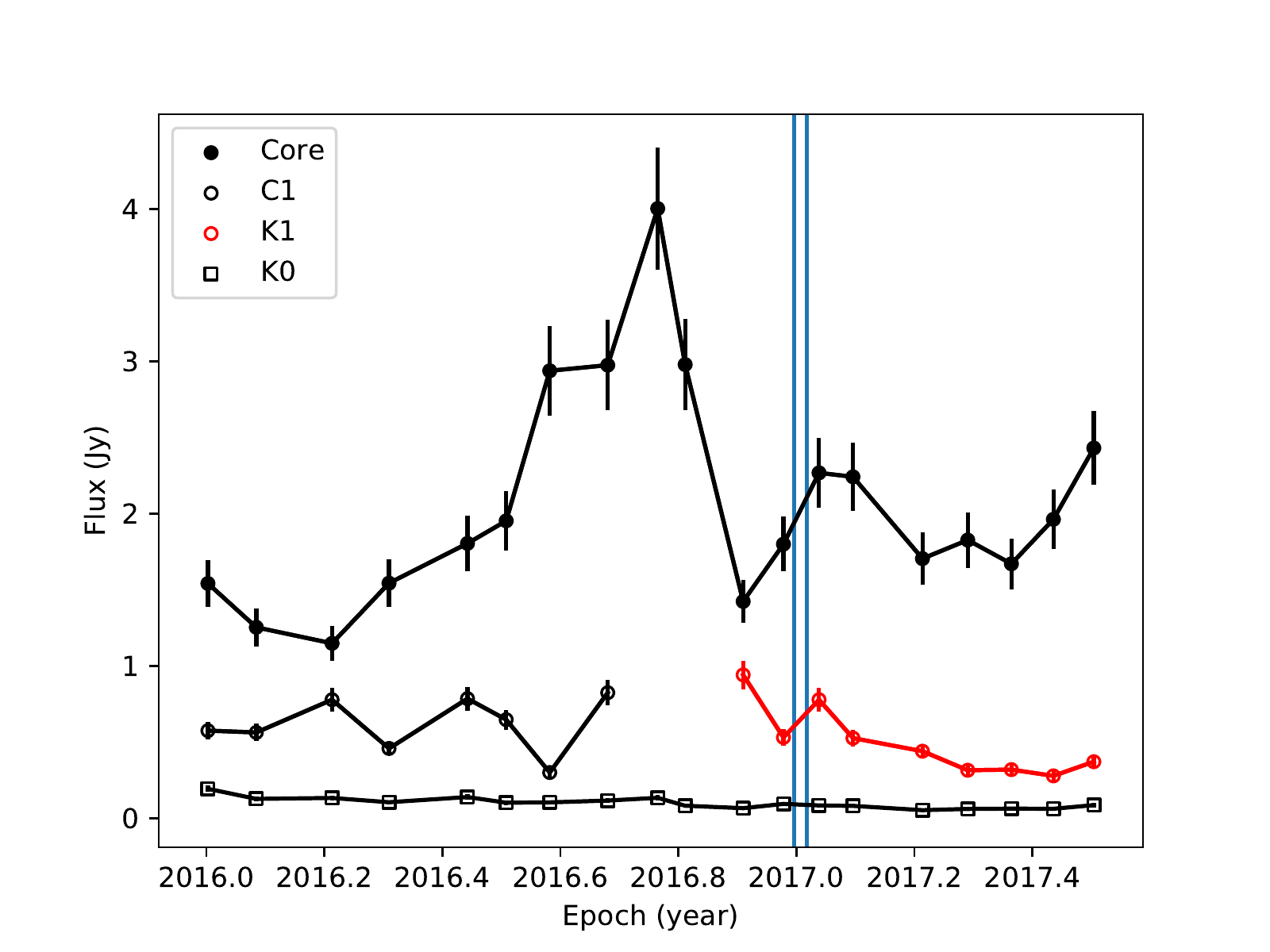}
   \caption{Light curves of 43 GHz VLBA model-fit components. Blue vertical lines are the same as in Fig.~\ref{Kinematic43}}
              \label{Flux43}%
    \end{figure}

The model-fit analysis at 43 GHz allowed us to investigate the kinematics and flux density variability of the radio jet during the flaring period; see Figs.~\ref{Kinematic43} and~\ref{Flux43}. We associated the core with the brightest unresolved component in the northwestern (upstream) end of the jet and we considered it to be stationary. Close to the core, at $\sim$0.1 mas, we detected another stationary component that we labeled C1, as we did in \cite{Casadio:2015fj}. Indeed, this stationary feature has already been observed in many previous studies \citep[e.g.,][]{2005AJ....130.1418J,2017ApJ...846...98J}, and was interpreted as a recollimation shock, which can trigger both radio \citep{2013A&A...551A..32F} and \g-ray outbursts \citep{Casadio:2015fj}. Another component, K0, is observed moving farther downstream in the jet.     


A new superluminal component, K1, has been visible in CTA~102 since November 2016 (see Fig.~\ref{Kinematic43}). From a linear fit of separation versus time we derived the velocity of K1, $\beta_{app}$ = 11.5$\pm$0.9$c$ (0.209$\pm$0.017 mas year$^{-1}$), and extrapolated the ejection time of the component (i.e., time of coincidence of the centroid of K1 with the centroid of the core), T$_{ej}$ = 2016.55$\pm$0.07 (18 July 2016). 
Considering the average angular sizes of K1 and the core, which are a = 0.08$\pm$0.01 mas and a$_{0}$ = 0.025$\pm$0.005 mas, respectively; the time K1 takes to exit the core is (a/2+a$_{0}$)/$\beta_{app}$ = 114 days. This means that K1 starts exiting from the core on 2016.85$\pm$0.04 (07 November 2016) and this coincides with a decrease in flux of the core which was previously in a flaring state; see Fig.~\ref{Flux43}.

Afterwards, at the beginning of 2017, both the light curves of K1 and the core show an increase in the flux density; the jet downstream, represented by K0,  instead remains constant in flux. This increase in flux of K1 coincides with its passage through the stationary feature C1 and with the brightest phase of the flaring period at gamma and X-ray frequencies (see Fig.~\ref{FigGamXray}), but also at optical and UV frequencies \citep{2017Natur.552..374R,2018ApJ...863..114G,2018A&A...617A..59K,2018ApJ...866...16P}. After that, component K1 restarts its decreasing trend moving downstream along the jet while the core continues to vary around the flux of 2~Jy for many months. 

In Fig.~\ref{Kinematic43} we also show the position of the 86 GHz model-fit components in the three available epochs. Aside from the core we detect two more components, with the component closer to the core matching with the position of C1 in the first epoch, while in the last two epochs it seems to be co-spatial with the position of the new emerging component. With the few available GMVA epochs and the large separation in time, it is difficult to obtain an unambiguous match between components. However, with the identification of components as above, the new superluminal feature, K1, would be visible first at 86 GHz, in September 2016, and only later on at 43 GHz, as expected because of opacity effects.    

Using the method in \cite{2005AJ....130.1418J} we estimate the Doppler factor associated with the flux variation observed in K1 when it crosses C1. The method relies on the comparison between the observed variability time scale, $\Delta t_{var}$, which is affected by the Doppler boosting, and the angular size of the component, which instead is not. The variability Doppler factor is hence derived as follows:
\begin{align}
\delta_{var}=\frac{s D_{L}}{c\Delta t_{var}(1+z)},
\end{align}
where {\it s} = 1.6{\it a} and {\it a} = FWHM of the Gaussian component during the epoch of maximum flux, and $D_{L}$ is the luminosity distance. The variability timescale is defined as $\triangle t_{\mathrm{var}}$=d{\it t}/ln($S_{\mathrm{max}}/S_{\mathrm{min}}$), where $S_{\mathrm{max}}$ and $S_{\mathrm{min}}$ are the measured maximum and minimum flux densities, respectively, and d{\it t} is the time in years between $S_{\mathrm{max}}$ and $S_{\mathrm{min}}$.
The only assumption here is that the variability time scale we observed corresponds to the light-travel time across the component. This occurs if the light crossing time is longer than the radiative cooling time and shorter than the adiabatic cooling time.

We wanted to test the hypothesis that the passage of K1 through the stationary feature C1 triggers particle acceleration and the multi-frequency outbursts in December 2016 - January 2017. We infer the Doppler factor from both the rising and decaying time of the K1 flux density when it crosses C1. Hence, for measuring $S_{\mathrm{max}}$, we considered the epoch of the peak in K1 when it is crossing the stationary feature (January 2017). $S_{\mathrm{min}}$ was measured in December 2016 and February 2017, respectively, for the rising and decaying variability time scale (see Fig.~\ref{Flux43}).   
The average variability Doppler factor we obtained considering the rising and decaying time of K1 crossing C1, is $\delta_{var}$ = 34$\pm$4. 

Subsequently, combining the variability Doppler factor with the estimated apparent velocity ($\beta_{app}$) we also obtained the viewing angle and the Lorentz factor associated with the variation [see \cite{2005AJ....130.1418J} and \cite{Casadio:2015fj} for the details]. The values we obtained are $\theta_{var}$ = 0.9$\pm$0.2$^{\circ}$ and $\Gamma_{var}$ = 20.9$\pm$1.9, respectively. 

\subsection{The polarized emission evolution at 43 GHz}

   \begin{figure*}[htpb]
   \centering
   \includegraphics[width=1.0\textwidth]{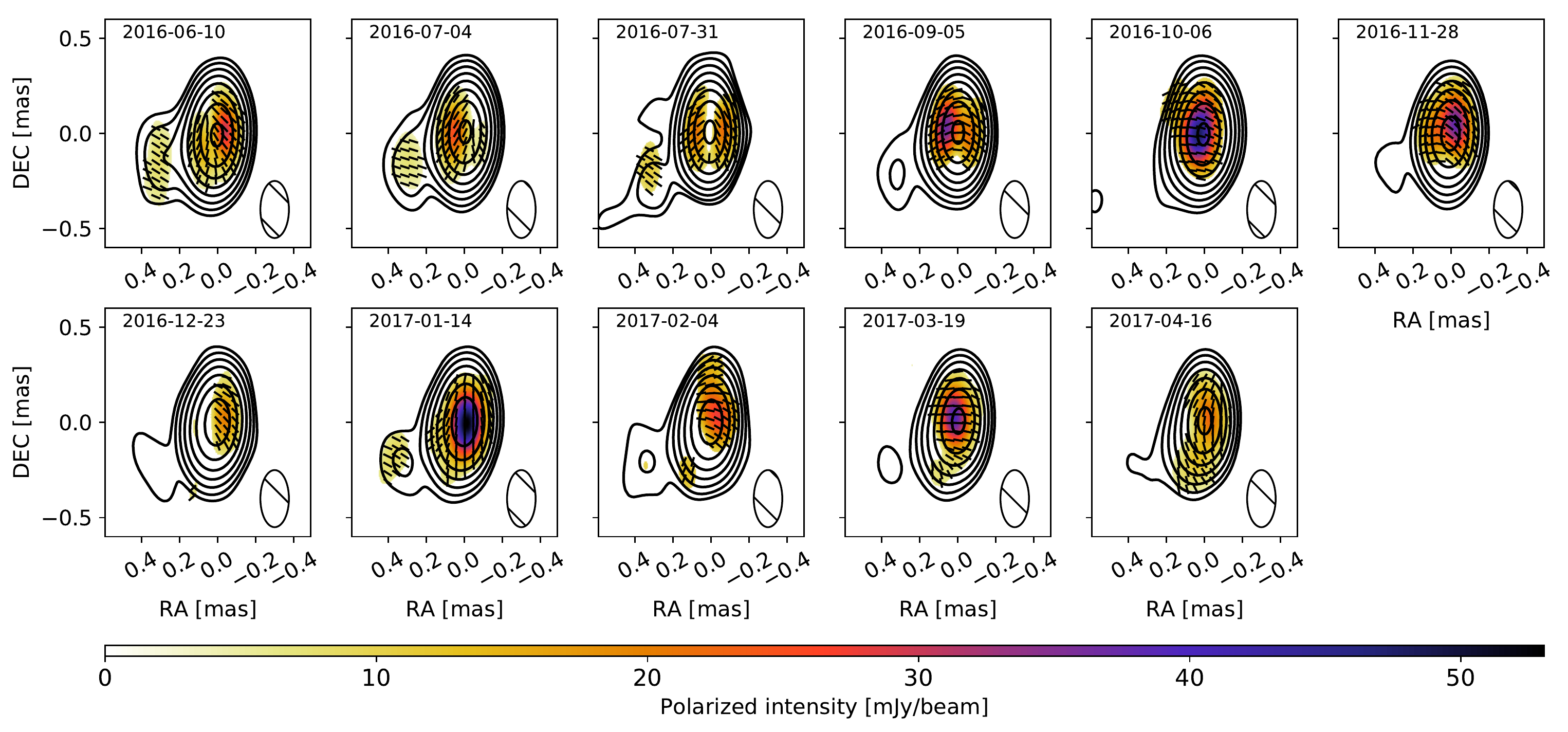}
   \caption{43 GHz VLBA polarimetric images of CTA~102. The common restoring beam is 0.15$\times$0.3 mas. Total intensity peaks are 2.35, 2.39, 3.09, 3.59, 3.56, 2.07, 351, 2.64, 2.44, 1.86, and 1.89 Jy/beam and contours are drawn at 0.8, 1.53, 3.02, 5.96, 11.74, 23.15, 45.65, and 90 $\%$ of 3.59 Jy/beam. Black sticks represent polarization vectors.}
              \label{43_1}%
    \end{figure*}
%

%
%

In Fig.~\ref{43_1} we display a series of 43 GHz total and linearly polarized intensity images that are used for the Faraday rotation analysis presented in Sect.~\ref{pol}. The images cover the period of the K1 ejection and exit from the core, as well as the crossing time through C1. The new superluminal component K1 is ejected from the core during July 2016 and it takes until the beginning of November to cross the core region. 

What we observe from Fig.~\ref{43_1} is that from July 2016 until September 2016, when K1 was crossing the core region, the EVPA orientation resembles the orientation present in the stacked image at 43 GHz (Fig.~\ref{concatenate}) and partially also the intrinsic orientation obtained from the Faraday rotation analysis (Fig.~\ref{RM}). The EVPAs in Fig.~\ref{RM} are slightly rotated after the correction for Faraday rotation. This tells us that during these two epochs we are observing the magnetic field structure in the inner jet which is highlighted by the passage of the superluminal component. In the following epoch, October 2016, the EVPAs are, in contrast, mainly oriented at 90$\degr$ and there is a peak in the polarized flux. We interpret this polarization morphology as a signature of the component K1 crossing the last optically thick surface of the core region and in turn highlighting the magnetic field structure present within this region of the jet.
Subsequently, November 2016, the component is already outside the core and the EVPAs remain partially horizontal, mainly in the region where the component is moving.

Later, we observe another peak in polarization in January and this time the EVPAs are mainly oriented at 0$\degr$. This is the epoch of the passage of K1 through C1 with K1 increasing its flux density. The fact that EVPAs here are oriented at 90$\degr$ with respect to their orientation in October 2016, and that C1 is still inside the 43 GHz beam, may suggests that the polarization vectors here are marking the magnetic field orientation in the stationary feature C1. 

However, we cannot discard the possibility that the polarized emission of this epoch is still associated with the core, which is also in a renewed high state of flux.  
A rotation of  90$\degr$ is expected because of opacity effects \citep{1970ranp.book.....P} or even because of changes of the jet orientation with respect to our line of sight \citep{2017MNRAS.467.3876L}. In both cases, however, a strong decrease of the degree of polarization is expected. Comparing the total intensity peaks and the polarized fluxes in the images of Fig.~\ref{43_1} we found no such decrease, leading us to discard both the aforementioned scenarios.

\section{Discussion}
In Sect.~\ref{pol} we presented the Faraday rotation analysis between 43 GHz VLBA and 86 GHz GMVA data, which reveals a RM gradient with a maximum value of $\sim$6$\times$10$^{4}$ rad/m$^{2}$ and a change of sign with position angle around the centroid of the core. The intrinsic EVPAs are observed rotating around the core. The RM gradient and EVPA orientation point to the presence of large-scale helical magnetic fields and a very small viewing angle  \citep{2010ApJ...725..750B,2011ApJ...737...42P}. This is also in agreement with the inferred viewing angle $\theta_{var}$ = 0.9$\pm$0.2$\degr$, obtained in Sect.~\ref{flares}. Considering this viewing angle and the projected distance of 3.86$\times$10$^{3}$ gravitational radii between the 86 GHz core and the black hole found in \cite{2015A&A...576A..43F}, we infer that the 86 GHz core is located at a de-projected distance of $\sim$2.5$\times$10$^{5}$ gravitational radii. Moreover, if we take into account the distance of $\sim$0.2 mas between the 15 and 86 GHz core regions reported in \cite{2015A&A...576A..43F}, the $\sim$7 mas distance from the 15 GHz core where \cite{Hovatta:2012fk} detected a significant RM gradient across the jet translates into $\sim$3.9$\times$10$^{7}$ gravitational radii. We demonstrate the existence of a helical magnetic field all the way from $\sim$2.5$\times$10$^{5}$ to $\sim$3.9$\times$10$^{7}$ gravitational radii.

The RM analysis is performed during a very active state of the source. Multi-wavelength flares are detected from December 2016 to January 2017. From our \g-ray and X-ray analysis we distinguish two main outbursts which occurred very close to one another in time (December 30 and January 07). Optical and UV flares were also detected very close in time to the high-energy flares \citep{2018A&A...617A..59K}.
The timing between the X-ray and \g-ray events suggests a common emitting region and this assumption allows us to infer a lower limit for the Doppler factor, $\delta_{\gamma}\gtrsim$15-17, needed to produce the observed high-energy emission.

From the 43 GHz kinematics analysis we find that a new superluminal component, K1, was ejected in July 2016 and moved downstream in the jet with an apparent speed of $\beta_{app}$ = 11.5$\pm$0.9$c$. The appearance of the component emerging from the core was accompanied by a decrease in the core flux density with the K1 light curve following a mainly decreasing trend attributable to adiabatic losses while the component propagates along the jet \citep{2013A&A...557A.105F}. However, this trend is interrupted by an increase in flux in January 2017, when the component K1 crosses a stationary feature (C1) which is located at $\sim$0.1 mas from the core. There is ample evidence \citep{2005AJ....130.1418J,2017ApJ...846...98J,2013A&A...551A..32F,2013A&A...557A.105F,Casadio:2015fj} that C1 is most likely associated with a standing recollimation shock in the jet.

The passage of K1 through the recollimation shock at 0.1 mas coincides with the flaring periods in the \g-ray, X-ray, UV, and optical pass bands. The Doppler factor that we deduce from the apparent velocity of K1 is around 11, but such a low value would not explain the observed high-energy emission, being incompatible with the lower limit for the Doppler factor we obtained in Sect.~\ref{flares}. If instead we infer the variability Doppler factor when the component K1 is crossing C1, we obtain a larger value: $\delta_{var}$ = 34$\pm$4. This value would explain the high-energy events and is also in agreement with the value found to explain the remarkable optical flare which occurred very close in time, on 28 December 2016 \citep{2017Natur.552..374R}. Moreover, a similar variability Doppler factor for CTA~102 is reported in \cite{Casadio:2015fj} and \cite{2017ApJ...846...98J}. 
In \cite{Casadio:2015fj} we found that a component with similar apparent speed ($\beta_{app}$ = 11.3$\pm$1.2$c$), slower than other previous components, was responsible for triggering an extraordinary multi-wavelength flare in 2012. The variability Doppler factor and viewing angle we inferred in that case were $\delta\sim$30 and $\theta\sim$1.2$\degr$.

The very small viewing angle obtained for K1 ($\theta_{var}$ = 0.9$\pm$0.2$\degr$) would explain the slowness of the component. We are not able to distinguish properly the motion of something moving toward us and this is the reason why the variability Doppler factor is more reliable than what we infer from the kinematics in these cases.

In Sect.~\ref{flares} we have also compared the motion of K1 along the jet with the evolution of the 43 GHz linearly polarized emission. During the passage of the component through the core region the polarization vector orientation resembles the intrinsic orientation obtained from the Faraday rotation analysis. In contrast, a different EVPA orientation is observed when the component is almost exiting the core (EVPAs mainly at 90$\degr$) and crossing the stationary feature C1 (EVPAs mainly at 0$\degr$). We hypothesise that this evolution in the orientation of the EVPAs may be attributed to the component highlighting the magnetic field structure within two distinct regions of the jet. The different orientation of the EVPAs, shown when the component K1 is crossing the last optically thick surface in the core region and when it crosses C1, could be evidence of a distinct magnetic field orientation in the two regions. We plan to investigate this scenario, comparing our observations with RMHD simulations in a forthcoming publication. A 90$\degr$ flip in the EVPAs due to opacity effects or different viewing angles is a less probable explanation considering the absence of an observed decrease in the level of fractional linear polarization.

%

\section{Conclusions}
We present mm-VLBI polarimetric images of the FSRQ CTA~102 with the highest possible resolution currently achievable ($\sim$50$\mu$as). The images were obtained with the GMVA at 86 GHz in May 2016, September 2016, and March 2017. Combining 43 GHz VLBA observations from the VLBA-BU-BLAZAR program with our 86 GHz GMVA data, we obtained the first high-resolution Faraday rotation image between these two frequencies. The Faraday rotation structure and the intrinsic EVPA orientation highlighted by the high-resolution image reveal the existence of a large-scale helical magnetic field in the very inner regions ($\sim$2.46$\times$10$^{5}$ gravitational radii, deprojected) of CTA~102 jet.   

From late 2016 to the beginning of 2017, CTA~102 was in a very active state, displaying extraordinary flares from radio to \g-ray frequencies. From the kinematics analysis at 43 GHz, we found that a new superluminal component was ejected from the mm core in July 2016. 
During the multi-wavelength outbursts, the new component was crossing another stationary feature located very close to the core ($\sim$0.1 mas), increasing its flux density. This supports the nature of recollimation shock of the stationary feature at $\sim$0.1 mas and the expected particle acceleration during shock-shock interaction. In our scenario, the interaction between the moving component and the recollimation shock triggers the multi-wavelength flares from December 2016 to January 2017. This is also supported by the variability Doppler factor associated with such an interaction ($\delta_{var}$ = 34$\pm$4), which is enough to explain the brightest \g-ray and X-ray flares observed in that period as well as the increase of 6-7 magnitudes in optical on 28 December 2016 \citep{2017Natur.552..374R}.

\begin{acknowledgements}
This research has made use of data obtained with the Global Millimeter VLBI Array (GMVA), which consists of telescopes operated by the MPIfR, IRAM, Onsala, Metsahovi, Yebes and the VLBA. The data were correlated at the correlator of the MPIfR in Bonn, Germany. The VLBA is an instrument of the National Radio Astronomy Observatory, a facility of the National Science Foundation of the USA operated under cooperative agreement by Associated Universities, Inc. (USA). The research at Boston University is supported by the NASA Fermi GI grant 80NSSC17K0649. Figure~\ref{pol_test}
also appears in \cite{2017Galax...5...67C}, published by Galaxies (an open access journal). We would like to thank the MPIfR internal referee Rocco Lico for the careful reading of the manuscript. We would also like to thank Helge Rottmann, Pablo de Vicente, Salvador Sanchez, Uwe Bach, Michael Lindqvist, and Jun Yang for the valuable GMVA data acquisition work.
This paper is partly based on observations carried out with the IRAM 30\,m Telescope. IRAM is supported by INSU/CNRS (France), MPG (Germany) and IGN (Spain). DB acknowledges support from the European Research Council (ERC) under the European Union’s Horizon 2020 research and innovation programme under grant agreement No 771282. IA acknowledges support by a Ram\'on y Cajal grant of the Ministerio de Ciencia, Innovaci\'on y Universidades (MICINU) of Spain. The research at the IAA–CSIC was supported in part by the MICINU through grant AYA2016-80889-P.
\end{acknowledgements}


\bibliographystyle{aa}
\bibliography{CTA102_GMVA}

\appendix
\section{$D$-terms calibration} \label{app:A}
In polarimetric VLBI observations the signal is recorded by two orthogonal feeds, right and left circular, or RCP and LCP. The signal recorded by each feed not only contains the information on the polarized signal coming from the source, but also part of the signal of the other receiver that "leaks" into it. The "leakages" or "$D$-terms" quantify the instrumental polarization which has to be removed from data before the final imaging.
The $D$-terms depend only on the station hardware and they are expected to change slowly with time \citep{Gomez:2002memo}. We used the task LPCAL in {\tt AIPS} to estimate the $D$-terms for all the sources observed under the aforementioned GMVA monitoring program (PI: A. Marscher) in the three epochs presented in this study. The LPCAL also provides the real parallactic angle coverage of each antenna (i.e., considering the scans kept after the calibration process), and this is an important parameter for the accuracy of the polarization calibration \citep{Leppanen:1995fv}. Hence, we expect to have the same $D$-terms for all the sources with a measurement accuracy that depends mostly on the  parallactic angle coverage, and in part also on the complexity of the polarized source substructure. 
\begin{figure*}[htpb]
   \centering
   \includegraphics[width=1.15\textwidth]{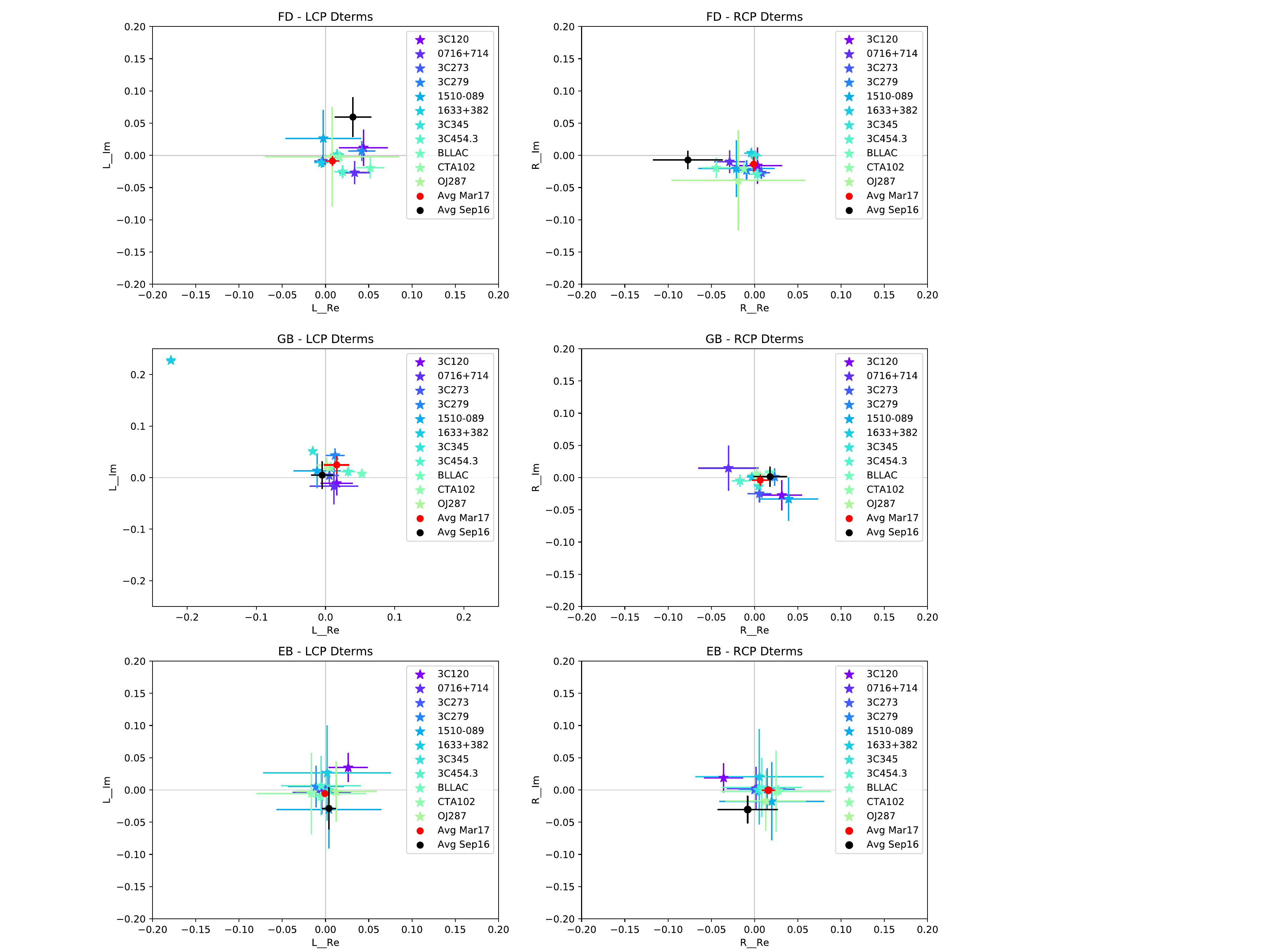}
   \caption{$D$-terms measurements obtained for all the sources observed in the 3 mm GMVA observation in March 2017 for FD, GB, and EB feeds. Red and black dots represent the weighted average values obtained for March 2017 and September 2016, respectively.}
              \label{dterms}%
\end{figure*}
In order to test the validity of $D$-term measurements, for each antenna feed (RCP and LCP) we compared the measurements coming from all the sources and we did this in all the three GMVA epochs, always taking LA as reference antenna. In Fig.~\ref{dterms} we report, as example, the plots obtained for three antennas for the epoch of September 2016. The task LPCAL unfortunately does not provide any uncertainties on measurements; therefore we decided to use the parallactic angle coverage as weights for the relative comparisons between measurements. The $D$-terms magnitudes are in the range 1$\%$ - 15$\%$, in agreement with previous works \cite{1994ApJ...427..718R,2012A&A...542A.107M}, but we also found some outliers, as for example 1633+382 in GB-LCP in September 2016 (see Fig.~\ref{dterms}), which we have not considered in our final analysis. Moreover, from the analysis of resulting plots we also decided to remove $D$-terms values with parallactic angle coverage below 70$\degr$. With the remaining measurements, we inferred weighted average values and the associated standard deviations for each feed and epoch. The standard deviations are around 1-3$\%$. 

The obtained average values for VLBA stations are not stable even on a six-monthly scale, contrary to what is reported in \cite{Gomez:2002memo}. The reason could be the 86 GHz receivers which do not provide stable $D$-terms or other hardware changes in VLBA stations. In any case, we cannot use the $D$-terms stability as a method for the calibration of the absolute orientations of EVPAs among epochs \citep{Leppanen:1995fv, Gomez:2002memo}.      

   \begin{figure*}[htpb]
   \centering
   \includegraphics[width=0.8\textwidth]{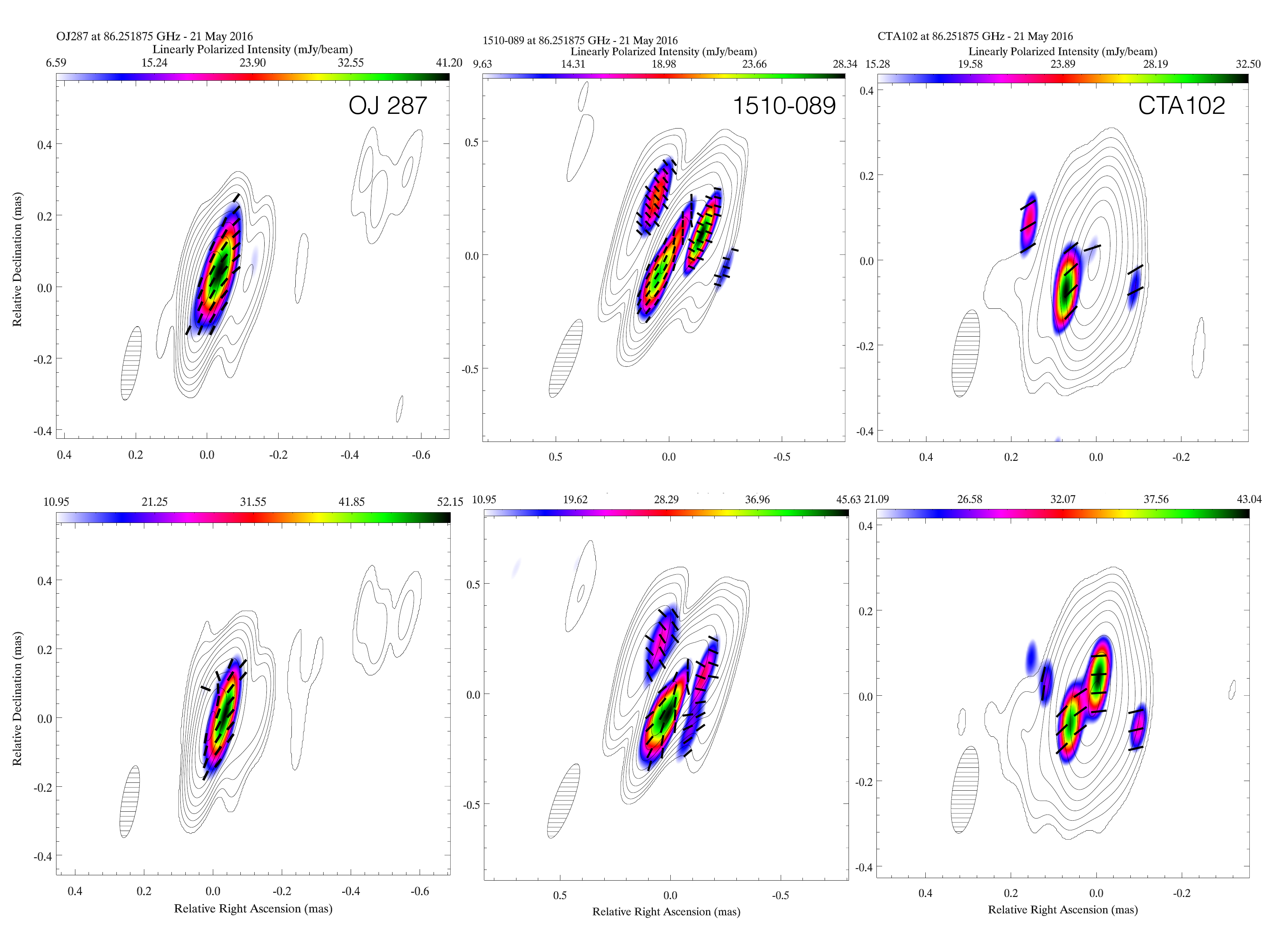}
   \caption{Series of 3 mm GMVA images in total (contours) and linearly polarized (colours) intensity of OJ~287, 1510-089, and CTA~102 taken on 21 May 2016. Black sticks represent the EVPAs. The images on top are obtained applying to the sources their own $D$-terms, while the bottom images result from the application of the average $D$-terms.}
              \label{pol_test}%
    \end{figure*}

We tested three different approaches for the calibration of the instrumental polarization. In the first epoch (May 2016) we chose three sources (OJ~287, 1510-089 and CTA~102) with different parallactic angle coverage and imaged each source using the $D$-terms obtained from its own data and the set of average $D$-terms obtained as described above. Figure~\ref{pol_test} displays the images of the three sources obtained applying both methods. In the case of OJ~287, which presents the best parallactic angle coverage (between 100$\degr$ and 120$\degr$ for most of the antennas), both methods give similar results; while for the other two sources, which have worse parallactic angle coverage (most of the antennas are below 95$\degr$) the resulting images present differences. The morphology of the polarized emission changes substantially between the two images, and the peak of the linearly polarized intensity shifts to a different position. Moreover, applying the average $D$-terms we obtained higher dynamic ranges in polarization confirming the influence of feed calibration errors in the dynamic range determination of the polarization images \citep{Leppanen:1995fv}.

\begin{figure*}[htpb]
   \centering
   \includegraphics[width=0.8\textwidth]{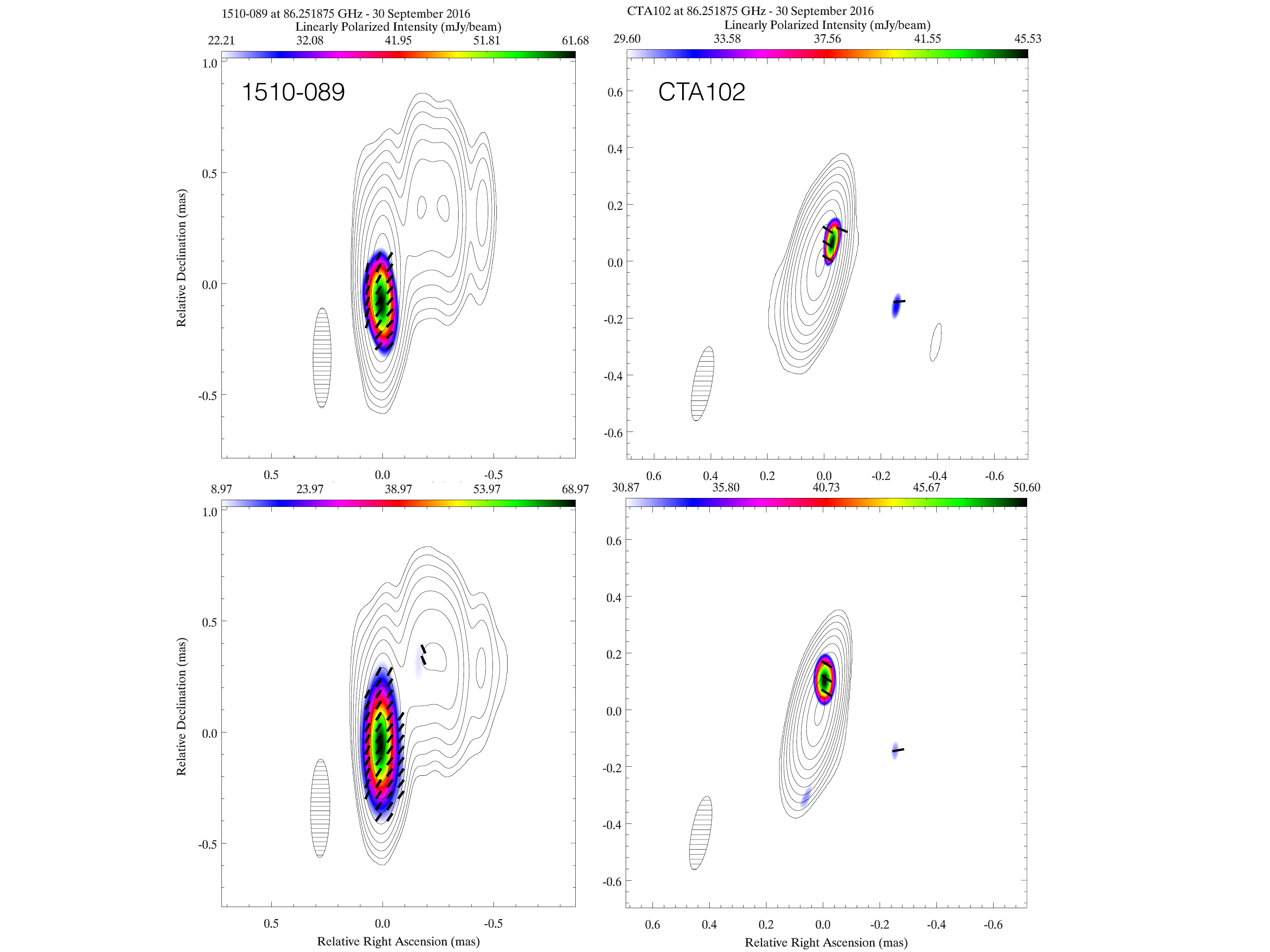}
   \caption{3 mm GMVA polarimetric images of 1510-089 and CTA~102 in September 2016. Contours and colors represent the same as in Fig.~\ref{pol_test}. The images on top are obtained applying the $D$-terms of 0716+714 and the bottom images applying the average $D$-terms.}
              \label{pol_test2}%
    \end{figure*}

The third method consists in obtaining the $D$-terms of only one source, compact and with good parallactic angle coverage, and applying them to all the other sources. We tested this approach in the epoch of September 2016, using the $D$-terms of 0716+714 which is a compact source and in that epoch has a very good parallactic angle coverage in all the antennas. Figure~\ref{pol_test2} displays the resulting images of 1510-089 and CTA~102 applying the $D$-terms of 0716+14 (top) and the average $D$-terms (bottom). Because of the good parallactic angle coverage of 0716+714, the images this time show no substantial difference, although applying the average $D$-terms we still have higher dynamic ranges. 

From the comparison of the three methods used for the calibration of the instrumental polarization, we decide to use the method of the average $D$-terms for all the sources and for the three GMVA epochs.  

\end{document}